\documentclass[aps,showpacs,pra,superscriptaddress,]{revtex4}
\usepackage{amsmath}
\usepackage{amsfonts}
\usepackage{amssymb}
\usepackage{bm}
\usepackage{graphicx}

\begin{document}

\title{Propagation of periodic and solitary waves in a highly dispersive
cubic-quintic medium with self-frequency shift and self-steepening
nonlinearity}
\author{Vladimir I. Kruglov}
\affiliation{Centre for Engineering Quantum Systems, School of Mathematics and Physics, The University of Queensland, Brisbane, Queensland 4072, Australia}
\author{ Houria Triki}
\affiliation{Radiation Physics Laboratory, Department of Physics, Faculty of Sciences, Badji Mokhtar University, P. O. Box 12, 23000 Annaba, Algeria}

\begin{abstract}
We study the dynamics of femtosecond light pulse propagation in a
cubic-quintic medium exhibiting dispersive effect up to the fourth order as well as self-frequency shift and self-steepening nonlinearity. A rich variety of periodic and solitary wave solutions are derived for the governing generalized higher-order nonlinear Schr\"{o}dinger equation in the presence of self-frequency shift and self-steepening effects. It is found that the frequency shift,
inverse velocity, amplitude and wave number of both periodic and solitary waves depend on dispersion coefficients and nonlinearity parameters as well. The conditions on optical fiber parameters for the existence of these structures are presented. The stability of these periodic and solitary wave solutions is studied numerically by adding white noise. It is proved by using the numerical split-step Fourier method that the profile of these nonlinear waves remains unchanged during evolution. 

\end{abstract}

\maketitle

\section{Introduction}

Solitons transmission in Kerr-type nonlinear media has been studied
extensively because of their important potential applications to all-optical switching and to high-capacity fiber optical communications \cite{R1,R2}. Such transmitting media is characterized by a refractive index which varies linearly with the light intensity. The main governing equation that arises for the description of picosecond light pulse propagation inside such optical system is the cubic nonlinear Schr\"{o}dinger (NLS) equation, which is usually derived with use of the slowly-varying envelope approximation.
This equation is known by its so-called soliton solutions which are formed due to the exact balance between group velocity dispersion and nonlinearily induced self-phase modulation. Such localized-in-time structures can propagate in optical fibers without changing their shape, which make them significant objects from both experimental and theoretical point of views. It is worthy to note that in addition to their observation in single mode fibers \cite{M}, soliton pulses have been also observed in other physical media such as femtosecond lasers \cite{S} and bulk optical materials \cite{A} in which the role of dispersion is replaced by spatial diffraction.

As peak powers increase and reduce the pulse duration the NLS
equation becomes not sufficient to ensure an accurate description of
ultrashort pulse propagation inside the fiber. Particularly, the inclusion of third-order dispersion term in the model is shown to be essential, especially for pulse durations below $10$ fs \cite{P1,P2}. As the light pulses become extremely short (below $6$ fs), additional higher-order effects such as self-steepening and self-frequency shift due to stimulated Raman scattering should also be taken into account. Experimentally, such shorter pulses with a duration of only $6$ fs was achieved by compressing amplified pulses from a colliding pulse mode-locked (CPM) dye laser \cite{Brito}. The quintic nonlinearity has been paid particular attention in recent years  \cite{T1,T2,T3}, which is featured by several optical materials such as semiconductor-doped glasses \cite{Roussignol}, chalcogenide glasses \cite{Chen}, dye solutions \cite{Ganeev}, and colloids \cite{Dutta}. Quintic nonlinearity is also found in nonlinear optical organic materials such as polydiacetylene para-toluene sulfonate \cite{Lawrence}. Such cubic-quintic nonlinear media have been studied as the key elements for future photonic and telecommunication technologies \cite{Liu,Serkin}. An interesting finding is that the cubic-quintic nonlinearity may lead to a formation of sufficiently robust optical vortices for cylindrical light beam propagation \cite{Quiroga}. In this context, the propagation of stable two-dimensional spatial solitons was recently reported in liquid CS$_{2}$ due to the simultaneous contributions of third- and fifth-order susceptibilities respectively \cite{Falcao}.

The theoretical description of ultrashort pulse propagation in an optical material exhibiting third- and fifth-order susceptibilities is based on the so-called cubic-quintic NLS equation that is established for the first time in Ref. \cite{Kh}. Various soliton solutions including bright and dark soliton solutions have been obtained for this model with both constant \cite{Zhang,Tanev} and variable coefficients \cite{Loomba}. As the pulse width becomes even narrower (below 10 fs), the fourth-order dispersion becomes also significant \cite{Palacio1,Palacio2}. In this situation, an extended higher-order NLS (HNLS) equation incorporating both cubic-quintic nonlinear
terms as well as third- and fourth-order dispersions is the main 
equation to describe the field dynamics in a medium that exhibits a
parabolic nonlinearity law \cite{Palacio1,Palacio2}. This equation has demonstrated its importance in nonlinear optics \cite{Tanev} and other
physical areas such as the coalescence of droplets in the first-order phase transition in Bose-Einstein condensates \cite{Bara} and the propagation of solitonic bubbles in a bosonic gases \cite{Jos}. In the present study, considering the influence of self-steepening and self-frequency shift effects on ultrashort pulse propagation, we study the existence of periodical and solitary waves in a cubic-quintic optical medium exhibiting all orders of dispersion up to the
fourth-order as well as self steepening and self-frequency shift effects. Remarkably, we find that the system supports a rich variety of different types of periodic and solitary waves in the presence of all higher-order effects.

The structure of this paper is as follows. First, in Sec. II, the
generalized HNLS equation modeling the evolution of femtosecond optical
pulses in a highly dispersive cubic-quintic media in presence of
self-steepening and self-frequency shift effects is presented. The nonlinear equation that governs the dynamics of wave amplitude in the system is also derived here. Then, in Sec. III, we present the exact periodic wave solutions of the mentioned equation and discuss the propagation properties and formation conditions of their existence. Results in the long wave limit are also discussed here, which leads to the formation of bright- and dark-type solitary waves. We also analyze the stability of nonlinear wave solutions by numerical simulation. Finally, the results are summarized in Sec. IV.

\section{Generalized HNLS equation and traveling waves}

The femtosecond pulse propagation through a highly dispersive cubic-quintic media is governed by the following generalized HNLS equation, 
\begin{equation}
i\frac{\partial \psi }{\partial z}=\alpha \frac{\partial ^{2}\psi }{\partial\tau ^{2}}+i\sigma \frac{\partial ^{3}\psi }{\partial \tau ^{3}}-\epsilon \frac{\partial ^{4}\psi }{\partial \tau ^{4}}-\gamma \left\vert \psi\right\vert ^{2}\psi +\mu \left\vert \psi \right\vert ^{4}\psi -i\rho \psi \frac{\partial }{\partial \tau }(\left\vert \psi \right\vert ^{2})-i\nu \frac{\partial }{\partial \tau }(\left\vert \psi \right\vert ^{2}\psi ),
\label{1}
\end{equation}%
where $\psi (z,\tau )$ is the complex field envelope, $z$ represents the
distance along direction of propagation, and $\tau $ is the retarded
time in the frame moving with the group velocity of wave packets. The (real) parameters $\alpha =\beta _{2}/2$, $\sigma =\beta _{3}/6$, and $\epsilon=\beta _{4}/24$ are related to second order dispersion $\beta _{2}$, third-order dispersion $\beta _{3}$, and fourth-order dispersion $\beta _{4}$ coefficients respectively.The parameters $\gamma $ and $\mu $ govern the effects of cubic and quintic nonlinearity respectively, while $\rho $ and $\nu $ represent the self-frequency shift and self-steepening coefficients respectively.

To obtain traveling wave solutions of the generalized HNLS equation (\ref{1}), we consider a solution of the form, 
\begin{equation}
\psi (z,\tau )=u(\xi )\exp [i(\kappa z-\delta \tau +\theta )],  \label{2}
\end{equation}%
where $u(\xi )$ is a real amplitude function which depends on the variable $\xi =\tau -qz$, were $q=\mathrm{v}^{-1}$ is the inverse velocity. The real parameters $\kappa $ and $\delta $ represent the wave number and frequency shift respectively, while $\theta $ represents the phase of pulse at $z=0$.

Substituting the traveling wave function (\ref{2}) into Eq. (\ref{1}), we
obtain the system of ordinary differential equations as 
\begin{equation}
(\sigma +4\epsilon \delta )\frac{d^{3}u}{d\xi ^{3}}+[s-(2\rho +3\nu )u^{2}]\frac{du}{d\xi }=0,  \label{3}
\end{equation}%
\begin{equation}
\epsilon \frac{d^{4}u}{d\xi ^{4}}-n\frac{d^{2}u}{d\xi^{2}}-mu+pu^{3}-\mu
u^{5}=0.~~~~~~  \label{4}
\end{equation}%
It is used here the following notations: 
\begin{equation}
n=\alpha +3\sigma \delta +6\epsilon \delta ^{2},~~~~m=\kappa -\alpha \delta^{2}-\sigma \delta ^{3}-\epsilon \delta ^{4},  \label{5}
\end{equation}%
\begin{equation}
p=\gamma +\nu \delta ,~~~~s=q-2\alpha \delta -3\sigma \delta ^{2}-4\epsilon\delta ^{3}.  \label{6}
\end{equation}%
Multiplying Eq. (\ref{4}) by the function $du/d\xi $ and integrating the
resulting equation, one arrives at the following nonlinear differential
equation, 
\begin{equation}
\epsilon \frac{du}{d\xi }\frac{d^{3}u}{d\xi^{3}}-\frac{\epsilon }{2}\left(\frac{d^{2}u}{d\xi^{2}}\right) ^{2}-\frac{n}{2}\left( \frac{du}{d\xi }\right)^{2}-\frac{m}{2}u^{2}+\frac{p}{4}u^{4}-\frac{\mu }{6}u^{6}=C_{0}, \label{7}
\end{equation}%
where $C_{0}$ is an integration constant. We introduce a new function $F(u)$ as 
\begin{equation}
F(u)=\left( \frac{du}{d\xi }\right) ^{2},~~~~~~~  \label{8}
\end{equation}%
which transforms Eq. (\ref{7}) to the ordinary nonlinear differential
equation, 
\begin{equation}
\epsilon F\,\frac{d^{2}F}{du^{2}}-\frac{\epsilon}{4}\left(\frac{dF}{du}%
\right) ^{2}-nF-mu^{2}+\frac{p}{2}u^{4}-\frac{\mu }{3}u^{6}=2C_{0}.
\label{9}
\end{equation}%
The nonlinear differential equation (\ref{9}) has the following special
polynomial solution,%
\begin{equation}
F(u)=a+bu^{2}+cu^{4}.  \label{10}
\end{equation}%
Note that this solution is obtained for an integration constant as 
$C_{0}=a(\epsilon b-n/2)$. The parameters in Eq. (\ref{10}) are given by
equations, 
\begin{equation}
8\epsilon c^{2}=\frac{1}{3}\mu ,~~~~10\epsilon bc=nc-\frac{1}{2}p,
\label{11}
\end{equation}%
\begin{equation}
12\epsilon ac=m+nb-\epsilon b^{2}.  \label{12}
\end{equation}

Thus, Eqs. (\ref{8}) and (\ref{10}) lead to the ordinary nonlinear
differential equation, 
\begin{equation}
\left( \frac{du}{d\xi }\right)^{2}=a+bu^{2}+cu^{4},  \label{13}
\end{equation}
where the coefficients $c$, $b$ and $a$ are given by Eqs. (\ref{11}) and (\ref{12}) as 
\begin{equation}
c=\pm\frac{1}{2}\sqrt{\frac{\mu}{6\epsilon}},~~~~b=\frac{1}{10\epsilon}%
(\alpha+3\sigma\delta+6\epsilon\delta^{2})\mp\frac{(\gamma+\nu\delta)}{%
10\epsilon}\sqrt{\frac{6\epsilon}{\mu}},  \label{14}
\end{equation}
\begin{equation}
a=\frac{1}{12\epsilon c}
[\kappa-\alpha\delta^{2}-\sigma\delta^{3}-\epsilon%
\delta^{4}+b(\alpha+3\sigma\delta+6\epsilon\delta^{2})-\epsilon b^{2}].
\label{15}
\end{equation}
Thus, the coefficients given by Eq. (\ref{14}) can have upper or lover sign which leads to different solutions of Eq. (\ref{13}). 

We emphases that Eq. (\ref{4}) is not equivalent to Eq. (\ref{13}). However, all solutions of Eq. (\ref{13}) are the solutions of Eq. (\ref{4}) as well. We note that Eq. (\ref{13}) leads to the following nonlinear differential equation, 
\begin{equation}
\frac{d^{3}u}{d\xi ^{3}}=(b+6cu^{2})\frac{du}{d\xi }.  \label{16}
\end{equation}%
The substitution of Eq. (\ref{16}) to (\ref{3}) yields the system of
algebraic equations, 
\begin{equation}
b(\sigma +4\epsilon \delta )+s=0,~~~~6c(\sigma +4\epsilon \delta )=2\rho
+3\nu .  \label{17}
\end{equation}%
Note that these algebraic equations are necessary and sufficient for
fulfillment of Eq. (\ref{3}) in the set of nontrivial ($u(\xi )\neq const$) solutions of Eq. (\ref{13}). These equations lead to the frequency shift $\delta $ and inverse velocity $q=v^{-1}$ as%
\begin{equation}
\delta =-\frac{\sigma }{4\epsilon }\pm \frac{(2\rho +3\nu )}{12\epsilon }\sqrt{\frac{6\epsilon }{\mu }},  \label{18}
\end{equation}%
\begin{equation}
q=2\alpha \delta +3\sigma \delta ^{2}+4\epsilon \delta ^{3}+\frac{1}{5\mu }(2\rho +3\nu )(\gamma +\nu \delta )\mp \frac{1}{30\epsilon }(2\rho +3\nu)(\alpha +3\sigma \delta +6\epsilon \delta ^{2})\sqrt{\frac{6\epsilon }{\mu }}.  \label{19}
\end{equation}%
It follows from Eq. (\ref{15}) that the wave number $\kappa $ is 
\begin{equation}
\kappa =\alpha \delta ^{2}+\sigma \delta ^{3}+\epsilon \delta ^{4}-b(\alpha+3\sigma \delta +6\epsilon \delta ^{2})+\epsilon b^{2}+12\epsilon ca. \label{20}
\end{equation}%
Thus, the wave number $\kappa $ is defined by parameter $a$ and vice versa. We present below a number of periodic (or elliptic) solutions of the model Eq. (\ref{1}) based on the nonlinear differential equation (\ref{13}). These closed form solutions are expressed in terms of Jacobi elliptic functions of modulus $k$. We further show that special limiting cases of these families include the bright and dark solitary wave solutions.

\section{Results and discussion}

In this section, we show the existence of a rich set of periodical and
solitary waves in the highly dispersive cubic-quintic medium with
self-frequency shift and self-steepening effects governed by Eq. (\ref{1}).

\subsection{Periodic wave solutions}

We first consider the transformation of Eq. (\ref{13}) based on new function $y(\xi )$ as 
\begin{equation}
u^{2}(\xi )=-\frac{1}{4c}y(\xi ).~~~~~~~  \label{21}
\end{equation}%
Thus, we have the nonlinear differential equation, 
\begin{equation}
\left( \frac{dy}{d\xi }\right) ^{2}=\sigma _{1}y+\sigma
_{2}y^{2}-y^{3},~~~~~~~  \label{22}
\end{equation}%
where $\sigma _{1}=-16ac$ and $\sigma _{2}=4b$. The polynomial $f(y)=\sigma_{1}y+\sigma _{2}y^{2}-y^{3}$ has tree roots as 
\begin{equation}
y_{0}=0,~~~~y_{\pm }=2\left( b\pm g\right) ,~~~~g=\sqrt{b^{2}-4ac}.
\label{23}
\end{equation}%
Hence, Eq. (\ref{22}) can be written in the form, 
\begin{equation}
\left( \frac{dy}{d\xi }\right) ^{2}=-y(y-y_{-})(y-y_{+}).~~~~~~~  \label{24}
\end{equation}%
Below we present the set of periodical wave solutions of the model (\ref{1}) based on Eq. (\ref{24}) or (\ref{13}) with the frequency shift $\delta $, inverse velocity $q=v^{-1}$ and wave number $\kappa $ given by Eqs. (\ref{18}-\ref{20}).

\begin{description}
\item[\textbf{1. Periodic}$\sqrt{A+B\mathrm{cn^{2}}}-$\textbf{waves}] 
\end{description}

We can order the roots of polynomial $f(y)$ as $y_{0}<y_{-}<y_{+}$. In this case Eqs. (\ref{21}) and (\ref{24}) yield the periodic solution as 
\begin{equation}
u(\xi )=\pm [A+B\mathrm{cn^{2}}(w(\xi -\xi _{0}),k)]^{1/2},~~~~~~~
\label{25}
\end{equation}%
where $0<k<1$. This solution yields the relation for modulus $k$ of Jacobi elliptic function $\mathrm{cn}(w(\xi -\xi _{0}),k)$ as $k=\sqrt{2g/(b+g)}$. Hence, we have the parameter $a$ as 
\begin{equation}
a=\frac{b^{2}(1-k^{2})}{c(2-k^{2})^{2}}.  \label{26}
\end{equation}%
The parameters of periodic solution given in Eq. (\ref{25}) are 
\begin{equation}
A=\frac{b(k^{2}-1)}{c(2-k^{2})},~~~~B=-\frac{bk^{2}}{c(2-k^{2})},~~~~~~~
\label{27}
\end{equation}%
\begin{equation}
w=\sqrt{\frac{b}{2-k^{2}}}.~~~~~~~  \label{28}
\end{equation}%
It follows from this solution the conditions for parameters as $c<0$ and $b>0 $. The wave number $\kappa $ by Eqs. (\ref{20}) and (\ref{26}) is 
\begin{equation}
\kappa =\alpha \delta ^{2}+\sigma \delta ^{3}+\epsilon \delta ^{4}-b(\alpha
+3\sigma \delta +6\epsilon \delta ^{2})+\epsilon b^{2}+\frac{12\epsilon
b^{2}(1-k^{2})}{(2-k^{2})^{2}}.  \label{29}
\end{equation}%
Substitution of the solution (\ref{25}) into the wave function (\ref{2})
yields the following family of periodic wave solutions for the generalized HNLS equation (\ref{1}): 
\begin{equation}
\psi (z,\tau )=\pm \lbrack A+B\mathrm{cn^{2}}(w(\xi -\xi _{0}),k)]^{1/2}\exp
[i(\kappa z-\delta \tau +\theta )],  \label{30}
\end{equation}%
where modulus $k$ is an arbitrary parameter in the interval $0<k<1$ and $\xi_{0}$ is an arbitrary real constant. We note that in the limiting cases with $k=1$ this periodic wave reduces to a bright-type soliton solution.

\begin{description}
\item[\textbf{2. Periodic} $\mathrm{cn-}$\textbf{waves}] 
\end{description}

We can order the roots of polynomial $f(y)$ as $y_{-}<y_{0}<y_{+}$. In this case Eqs. (\ref{21}) and (\ref{24}) yield the periodic solution as%
\begin{equation}
u(\xi )=\pm \Lambda \mathrm{cn}(w(\xi -\xi _{0}),k),  \label{31}
\end{equation}%
where $k$ is an arbitrary parameter. This solution yields the relation for modulus $k$ of Jacobi elliptic function $\mathrm{cn}(w(\xi -\xi _{0}),k)$ as $k=\sqrt{(b+g)/2g}$. Hence, we have the parameter $a$ as%
\begin{equation}
a=\frac{b^{2}k^{2}(k^{2}-1)}{c(2k^{2}-1)^{2}}.  \label{32}
\end{equation}%
The parameters $\Lambda $ and $w$ are given by%
\begin{equation}
\Lambda=\sqrt{\frac{-bk^{2}}{c(2k^{2}-1)}},~~~~w=\sqrt{\frac{b}{2k^{2}-1}}. \label{33}
\end{equation}%
In this solution the modulus $k$ can belong two different intervals: $0<k<1/\sqrt{2}$ or $1/\sqrt{2}<k<1$. It follows from this solution the conditions for parameters as $c<0$ and $b<0$ when $0<k<1/\sqrt{2}$, and the conditions for parameters are $c<0$ and $b>0$ when $1/\sqrt{2}<k<1$. Thus, the wave number $\kappa $ by Eqs. (\ref{20}) and (\ref{32}) is%
\begin{equation}
\kappa =\alpha \delta ^{2}+\sigma \delta ^{3}+\epsilon \delta ^{4}-b(\alpha+3\sigma \delta +6\epsilon \delta ^{2})+\epsilon b^{2}+\frac{12\epsilon b^{2}k^{2}(k^{2}-1)}{(2k^{2}-1)^{2}}.  \label{34}
\end{equation}%
Substitution of the solution (\ref{31}) into the wave function (\ref{2})
yields the following family of periodic wave solutions for the generalized HNLS equation (\ref{1}):%
\begin{equation}
\psi (z,\tau )=\pm \Lambda \mathrm{cn}(w(\xi -\xi _{0}),k)\exp [i(\kappa
z-\delta \tau +\theta )],  \label{35}
\end{equation}%
where modulus $k$ is an arbitrary parameter in the interval $0<k<1/\sqrt{2}$ or $1/\sqrt{2}<k<1$. In the limiting case with $k=1$ this solution reduces to soliton solution.

\begin{description}
\item[\textbf{3. Periodic} $\mathrm{sn-}$\textbf{waves}] 
\end{description}

We can order the roots of polynomial $f(y)$ as $y_{-}<y_{+}<y_{0}$. In this case Eqs. (\ref{21}) and (\ref{24}) yield the periodic solution as%
\begin{equation}
u(\xi )=\pm \Lambda \mathrm{sn}(w(\xi -\xi _{0}),k),  \label{36}
\end{equation}%
where $0<k<1$. This solution yields the relation for modulus $k$ of Jacobi elliptic function $\mathrm{sn}(w(\xi -\xi _{0}),k)$ as $k=\sqrt{(b+g)/(b-g)}$. Hence, the parameter $a$ is given by
\begin{equation}
a=\frac{b^{2}k^{2}}{c(1+k^{2})^{2}}.  \label{37}
\end{equation}%
The parameters $\Lambda $ and $w$ are given by%
\begin{equation}
\Lambda=\sqrt{\frac{-bk^{2}}{c(1+k^{2})}},~~~~w=\sqrt{\frac{-b}{1+k^{2}}}. \label{38}
\end{equation}%
It follows from this solution the conditions for parameters as $c>0$ and $b<0 $. Thus, the wave number $\kappa $ by Eqs. (\ref{20}) and (\ref{37}) is%
\begin{equation}
\kappa =\alpha \delta ^{2}+\sigma \delta ^{3}+\epsilon \delta ^{4}-b(\alpha+3\sigma \delta +6\epsilon \delta ^{2})+\epsilon b^{2}+\frac{12\epsilon b^{2}k^{2}}{(1+k^{2})^{2}}.  \label{39}
\end{equation}%
Substitution of the solution (\ref{36}) into the wave function (\ref{2})
yields the following family of periodic wave solutions for the generalized HNLS equation (\ref{1}):%
\begin{equation}
\psi (z,\tau )=\pm \Lambda \mathrm{sn}(w(\xi -\xi _{0}),k)\exp [i(\kappa
z-\delta \tau +\theta )],  \label{40}
\end{equation}%
where modulus $k$ is an arbitrary parameter in the interval $0<k<1$. In the limiting case with $k=1$ this solution reduces to the kink wave solution.

\begin{description}
\item[\textbf{4. Periodic} $\mathrm{sn/}(1+\mathrm{dn})\mathrm{-}$\textbf{waves}] 
\end{description}

We have also found the periodic bounded solution of Eq. (\ref{13}) of the form,%
\begin{equation}
u(\xi )=\pm \frac{A\mathrm{sn}\left( w(\xi -\xi _{0}),k\right) }{1+\mathrm{dn}\left( w(\xi -\xi _{0}),k\right) },  \label{41}
\end{equation}%
where $0<k<1$, and the parameter $a$ in this solution is%
\begin{equation}
a=\frac{b^{2}k^{4}}{4c(2-k^{2})^{2}}.  \label{42}
\end{equation}%
The parameters $A$ and $w$ for this periodic solution are%
\begin{equation}
A=\sqrt{\frac{-bk^{4}}{2c(2-k^{2})}},~~~~w=\sqrt{\frac{-2b}{2-k^{2}}},
\label{43}
\end{equation}%
where $b<0$ and $c>0$. The wave number $\kappa $ by Eqs. (\ref{20}) and (\ref{42}) is%
\begin{equation}
\kappa =\alpha \delta ^{2}+\sigma \delta ^{3}+\epsilon \delta ^{4}-b(\alpha+3\sigma \delta +6\epsilon \delta ^{2})+\epsilon b^{2}+\frac{3\epsilon b^{2}k^{4}}{(2-k^{2})^{2}}.  \label{44}
\end{equation}%
Thus, Eq. (\ref{41}) yields the periodic bounded solution of Eq. (\ref{1}) as%
\begin{equation}
\psi (z,\tau )=\pm \frac{A\mathrm{sn}(w(\xi -\xi_{0}),k)}{1+\mathrm{dn}%
(w(\xi -\xi _{0}),k)}\exp [i(\kappa z-\delta \tau +\theta )],  \label{45}
\end{equation}%
where modulus $k$ is an arbitrary parameter in the interval $0<k<1$.

\begin{figure}[h]
\includegraphics[width=1\textwidth]{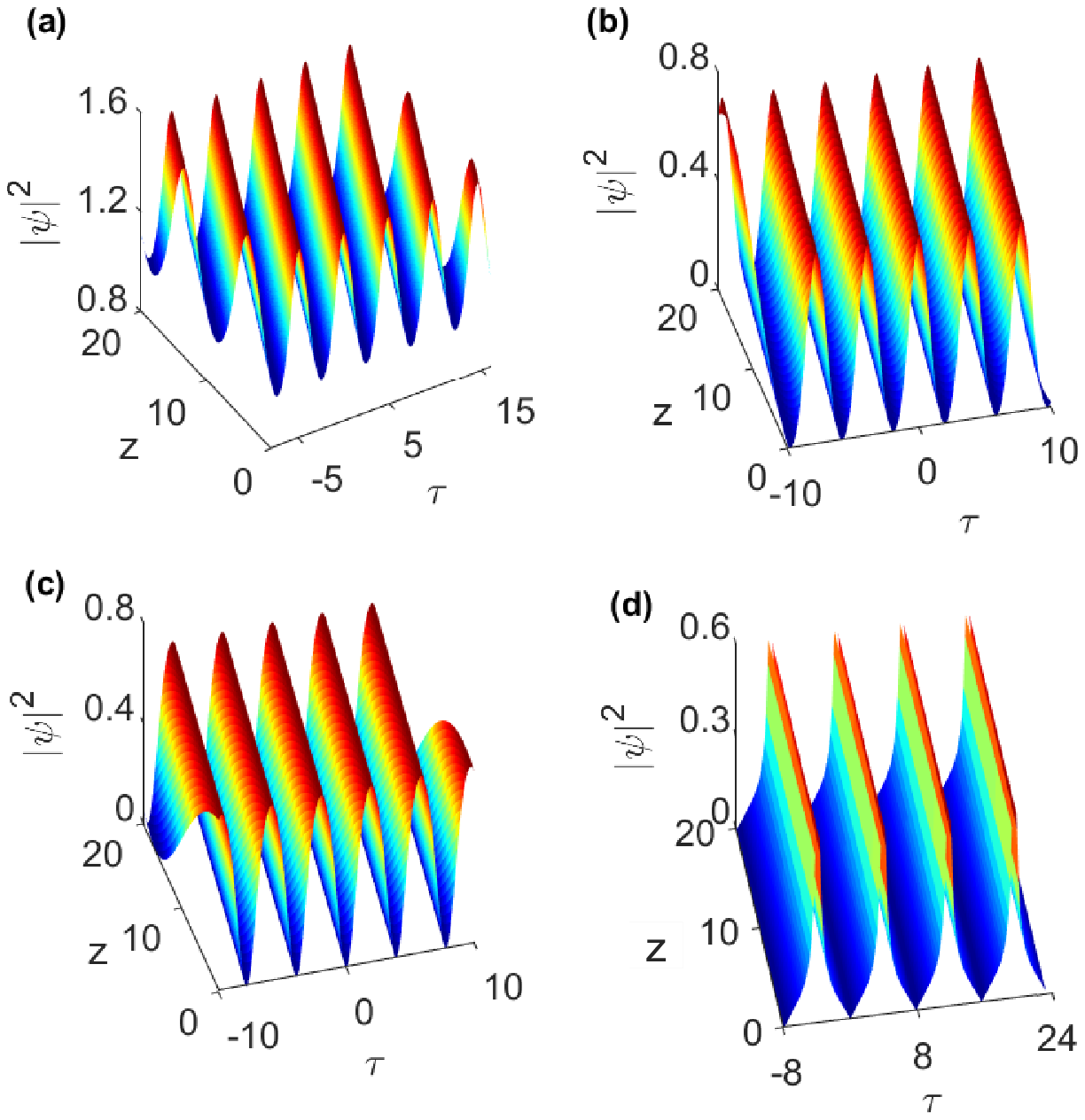}
\caption{Evolution of nonlinear wave solutions: (a) periodic $\mathrm{%
cn^{2}-}$wave (\ref{30}) (b) periodic $\mathrm{cn-}$wave (\ref{35}) (c) periodic $\mathrm{sn-}$wave (\ref{40}) and (d) periodic $\mathrm{sn/}\left( 1+\mathrm{dn}\right) \mathrm{-}$wave (\ref{45}) for the values mentioned in the text.}
\label{FIG.1.}
\end{figure}

The intensity profile of the periodic $\mathrm{cn^{2}-}$wave solution (\ref{30}) of Eq. (\ref{1}) is shown in Fig. 1(a) for the parameters values: $\alpha =0.5,$ $\sigma =-\frac{1}{3},$ $\epsilon =0.25,$ $\gamma =1.6,$ $\mu=1.5,$ $\rho =-0.25,$ and $\nu =-0.5$. The results for the periodic $\mathrm{cn-}$wave solution (\ref{35}) are presented in Fig. 1(b) for the values $\alpha =0.25,$ $\sigma =-\frac{1}{3},$ $\epsilon =0.25,$ $\gamma =1.75,$ $\mu =1.5,$ $\rho =-0.25,$ and $\nu =-0.5.$ The intensity profile of the periodic $\mathrm{sn-}$wave solution (\ref{40}) of Eq. (\ref{1}) is illustrated in Fig. 1(c) for the values $\alpha =0.2,$ $\sigma =-\frac{1}{3},$ $\epsilon =0.25,$ $\gamma =-0.925,$ $\mu =1.5,$ $\rho =0.25,$ and $\nu=0.5.$ Additionally, the intensity profile of the periodic $\mathrm{sn/}%
\left( 1+\mathrm{dn}\right) \mathrm{-}$wave solution (\ref{45}) is depicted in Fig. 1(d) for the parameters values: $\alpha =1,$ $\sigma =-\frac{1}{3},$ $\epsilon =0.25,$ $\gamma =1.05,$ $\mu =1.5,$ $\rho =0.25,$ and $\nu =0.5$.
Here the value of the elliptic modulus $k$ is taken as $k=0.6$ and  $\xi _{0}=0$. It can be seen from these figures that all the intensity profiles present the periodic property which makes them a model of optical pulse train propagation in nonlinear fiber systems. Unlike in the case of  periodic $\mathrm{cn}$\textrm{-}$\mathrm{,}$  $\mathrm{sn}$\textrm{-} and $\mathrm{sn/}\left( 1+\mathrm{dn}\right) $\textrm{-}wave solutions, the periodic wave of $\sqrt{A+B\mathrm{cn^{2}}}$-type is located on a nonzero background.

\subsection{\textbf{Solitary} wave solutions}

Studying the long wave limit ($k\rightarrow 1$) of periodical solutions is of practical significance as it leads to localized pulse profiles which are described by an hyperbolic secant or an hyperbolic tangent functions. Below we examine the formations and properties of bright and dark solitary waves in the optical medium by taking the $k\rightarrow 1$ limit in the solutions given in the preceding section.

\begin{description}
\item[\textbf{1. Bright solitary} $\mathrm{sech}$\textbf{$-$waves}] 
\end{description}

First, we consider the limiting case of solution in Eq. (\ref{25}) with $k=1$. Hence, we have the soliton solution of Eqs. (\ref{21}) and (\ref{24}) as%
\begin{equation}
u(\xi )=\pm \left( -\frac{b}{c}\right) ^{1/2}\mathrm{sech}(\sqrt{b}(\xi -\xi_{0})),  \label{46}
\end{equation}%
The condition $k=1$ in Eq. (\ref{29}) leads to the wave number $\kappa$ as%
\begin{equation}
\kappa =\alpha \delta ^{2}+\sigma \delta ^{3}+\epsilon \delta ^{4}-b(\alpha+3\sigma \delta +6\epsilon \delta ^{2})+\epsilon b^{2}.  \label{47}
\end{equation}%
Thus, using Eqs. (\ref{2}) and (\ref{46}) we have the bright solitary wave solution for generalized HNLS equation (\ref{1}) as%
\begin{equation}
\psi (z,\tau)=\pm\left(-\frac{b}{c}\right)^{1/2}~\mathrm{sech}(\sqrt{b}%
(\xi -\xi _{0}))\exp [i(\kappa z-\delta \tau +\theta )].  \label{48}
\end{equation}
This solitary wave exists for $b>0$ and $c<0$.

\begin{description}
\item[\textbf{2. Dark solitary} $\mathrm{tanh}$\textbf{$-$waves}] 
\end{description}

The limiting case with $k=1$ leads solution in Eq. (\ref{36}) to the kink wave solution as%
\begin{equation}
u(\xi )=\pm \Lambda _{0}\mathrm{tanh}(w_{0}(\xi -\xi _{0})).  \label{49}
\end{equation}%
The parameters of this solution are%
\begin{equation}
\Lambda_{0}=\left(-\frac{b}{2c}\right)^{1/2},~~~~w_{0}=\sqrt{\frac{-b}{2}},  \label{50}
\end{equation}%
where $b<0$ and $c>0$. The wave number $\kappa $ by Eq. (\ref{39}) and $k=1$ is%
\begin{equation}
\kappa =\alpha \delta ^{2}+\sigma \delta ^{3}+\epsilon \delta ^{4}-b(\alpha+3\sigma \delta +6\epsilon \delta ^{2})+4\epsilon b^{2}.  \label{51}
\end{equation}%
Hence, the kink solution for Eq. (\ref{1}) is given by 
\begin{equation}
\psi (z,\tau )=\pm \Lambda _{0}\mathrm{tanh}(w_{0}(\xi -\xi _{0}))\exp
[i(\kappa z-\delta \tau +\theta )].  \label{52}
\end{equation}%
Note that this kink solution has the form of dark soliton for intensity $I=|\psi (z,\tau )|^{2}=\Lambda _{0}^{2}\mathrm{tanh}^{2}(w_{0}(\xi -\xi
_{0}))$.

\begin{description}
\item[\textbf{3. Dark solitary} $\mathrm{tanh}/(1+\mathrm{sech})-$\textbf{waves}] 
\end{description}

The limit $k\rightarrow 1$ in Eq. (\ref{45}) leads to a solitary wave of the form, 
\begin{equation}
\psi (z,\tau )=\pm \frac{A_{0}\mathrm{tanh}\left( w_{0}(\xi -\xi
_{0})\right) }{1+\mathrm{sech}\left( w_{0}(\xi -\xi _{0})\right) }\exp
[i(\kappa z-\delta \tau +\theta )].  \label{53}
\end{equation}%
The parameters for this solitary wave are 
\begin{equation}
A_{0}=\sqrt{-\frac{b}{2c}},\qquad w_{0}=\sqrt{-2b},\qquad  \label{54}
\end{equation}%
with $b<0$ and $c>0$ and $a=b^{2}/4c$. This solitary wave has the form of dark soliton for intensity $I=|\psi (z,\tau )|^{2}$. The wave number $\kappa$ for this solitary wave follows from Eq. (\ref{44}) with $k=1$ as 
\begin{equation}
\kappa =\alpha \delta ^{2}+\sigma \delta ^{3}+\epsilon \delta ^{4}-b(\alpha
+3\sigma \delta +6\epsilon \delta ^{2})+4\epsilon b^{2}.  \label{55}
\end{equation}%
Remarkably, the functional form of the solitary wave (\ref{53}) differs from the simplest dark solitary \textrm{tanh}-wave.

\begin{figure}[h]
\includegraphics[width=1\textwidth]{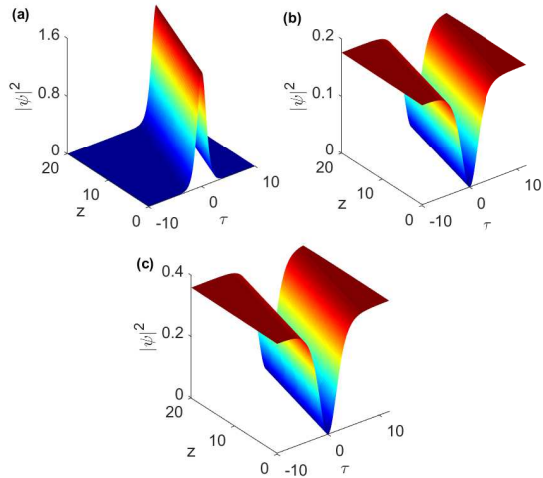}
\caption{Evolution of nonlinear wave solutions: (a) bright solitary
wave (\ref{48}) (b) dark solitary wave (\ref{52}) and (c) dark solitary wave (\ref{53}) for the values mentioned in the text.}
\label{FIG.2.}
\end{figure}

Figure 2(a) presents the evolution of the intensity profile of the bright solitary wave solution (\ref{48}) for the parameter values: $\alpha =0.5,$ $\sigma =-\frac{1}{3},$ $\epsilon =0.25,$ $\gamma =1.5,$ $\mu =1.5,$ $\rho=-0.25,$ and $\nu =-0.5$. The intensity profiles of the dark solitary wave solution (\ref{52}) and (\ref{53}) are shown in Fig. 2(b) and 2(c) respectively, using the parameter values: $\alpha =0.1,$ $\sigma =-\frac{1}{3},$ $\epsilon =0.25,$ $\gamma =1,$ $\mu =1.5,$ $\rho =0.25,$ and $\nu =0.5.$
We notice that these localized structures exist in the optical medium due to the balance between cubic and quintic nonlinearities, all orders of dispersion up to the fourth order, self-frequency shift, and self-steepening effects.

From the above results, we see that the inverse velocity, the frequency
shift, the wave number and the amplitude of both periodic and localized
waves depend on the nonlinearity parameters, unlike the parameters of
nonlinear waves in absence of self-frequency shift and self-steepening
effects \cite{K1,Krug,K2}, which has a dependence on dispersion coefficients solely. Therefore, we can select these nonlinear parameters to control the formation of propagating waves in the fiber medium.

\begin{figure}[h]
\includegraphics[width=1\textwidth]{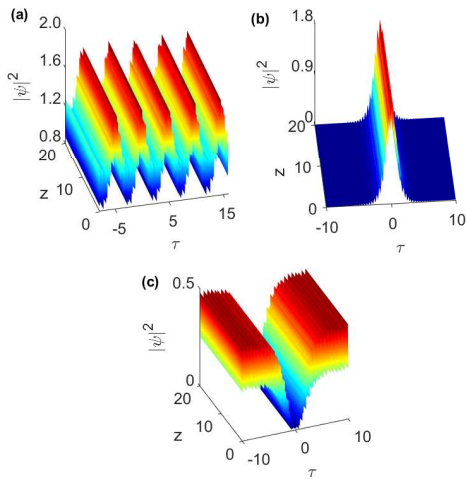}
\caption{Numerical evolution of the intensity $I=\left\vert \psi
\right\vert ^{2}$:  (a) the periodic $\mathrm{cn^{2}-}$wave solution (\ref{30}) (b) the bright solitary wave solution (\ref{48}) and (c) the dark solitary wave solution (\ref{53}) under the perturbation of white noise whose maximal value is $0.1$. The parameters are the same as in Figs. 1(a), 2(a) and 2(c) respectively.}
\label{FIG.3.}
\end{figure}

For the sake of completeness, we now analyze the stability of the obtained periodic and localized wave solutions with respect to finite perturbations. Here we take as examples the periodic wave (\ref{30}), bright solitary wave (\ref{48}) and dark solitary wave (\ref{53}) and demonstrate the stability of these solutions under the perturbation of the additive white noise. Thus, we perform numerical simulation using the split-step Fourier method \cite{Agraw}. As usual, we put the noise onto the initial profile, then the perturbed pulse reads \cite{JD}: $\psi _{\text{pert}}=\psi (\tau ,0)[1+0.1\,$ random$(\tau )].$ Figs. 3(a), 3(b) and 3(c) depict the evolution of nonlinear wave solutions (\ref{30}), (\ref{48}) and (\ref{53}) under the perturbation of $10\%$ white noise respectively. From these figure, we observe that the shape of the periodic and solitary waves remain unchanged while evolving along the propagation distance. Although we have shown here the results of stability study only for three examples of the HNLS model (\ref{1}), similar conclusions hold for other solutions as well. Therefore we
can conclude that the obtained nonlinear waves are stable and should be
observable in highly dispersive cubic-quintic optical media exhibiting
self-frequency shift and self-steepening processes.

\section{Conclusion}

We have studied the femtosecond light pulse propagation in a highly
dispersive optical fiber medium that is governed by a higher-order nonlinear Schr\"{o}dinger equation incorporating cubic and quintic nonlinearities, self-frequency shift, self-steepening, and all orders of dispersion up to the fourth order. With use of a regular method, we have derived the exact periodic wave solutions of the model in the presence of all higher-order dispersive and nonlinear effects. Solitary wave solutions have been also obtained in the long wave limit, which includes both bright and dark localized wave solutions. It is found that the physical parameters of these structures are dependent on the nonlinearity parameters. Moreover, we have checked the stability of the periodic and solitary wave solutions by using the numerical split-step Fourier method and the results have shown that the obtained nonlinear wave solutions are stable while evolving in distance. It is apparent that the exact nature of the nonlinear waves presented here can
lead to different applications in optical communications.

\end{document}